\documentclass[twocolumn]{aastex701}

\begin{document}

\title{Predictions for an unseen third planet in the WISPIT 2 system}

\author[0009-0008-3992-1890]{Amena Faruqi}
\email[show]{amena.faruqi@warwick.ac.uk}
\affiliation{Department of Physics, University of Warwick, Coventry CV4 7AL, UK}
\affiliation{Centre for Exoplanets and Habitability, University of Warwick, Coventry CV4 7AL, UK}

\author[0000-0002-3984-9496]{Farzana Meru}
\email{F.Meru@warwick.ac.uk}
\affiliation{Department of Physics, University of Warwick, Coventry CV4 7AL, UK}
\affiliation{Centre for Exoplanets and Habitability, University of Warwick, Coventry CV4 7AL, UK}

\author[0000-0002-2919-7500]{Gabriel-Dominique Marleau}
\email{gabriel.marleau@uni-due.de}
\affiliation{Fakult\"at f\"ur Physik,
Universit\"at Duisburg--Essen,
Lotharstra\ss{}e 1,
47057 Duisburg, Germany
}
\affiliation{Max-Planck-Institut f\"ur Astronomie,
K\"onigstuhl 17,
69117 Heidelberg, Germany
}
\affiliation{Division of Space Research and Planetary Sciences,
Physics Institute, University of Bern,
Gesellschaftsstr.~6,
3012 Bern, Switzerland}
\email{gabriel.marleau@uni-due.de}

\author[0000-0002-4448-3871]{Pooneh Nazari}
\email{Pooneh.Nazari@warwick.ac.uk}
\affiliation{Department of Physics, University of Warwick, Coventry CV4 7AL, UK}
\affiliation{Centre for Exoplanets and Habitability, University of Warwick, Coventry CV4 7AL, UK}

\author[0000-0001-6515-9854]{Andrew Swan}
\email{Andrew.Swan@warwick.ac.uk}
\affiliation{Department of Physics, University of Warwick, Coventry CV4 7AL, UK}


\author[0000-0002-7605-2961]{Ralph E. Pudritz}
\email{pudritz@mcmaster.ca}
\affiliation{Department of Physics and Astronomy, McMaster University, Hamilton, ON L8S 4K1, Canada}
\affiliation{Origins Institute, McMaster University, Hamilton, ON L8S 4K1, Canada}

\author[0000-0001-8749-1962]{Thomas G. Wilson}
\email{Thomas.G.Wilson@warwick.ac.uk}
\affiliation{Department of Physics, University of Warwick, Coventry CV4 7AL, UK}
\affiliation{Centre for Exoplanets and Habitability, University of Warwick, Coventry CV4 7AL, UK}

\begin{abstract}
We predict the mass and semi-major axis of a currently unseen but hypothesised planet in the WISPIT~2 system based on the location of the ring observed by ALMA at 144.4~au. Using findings from previous hydrodynamical models, we predict that a $\sim 0.2$--$1~M_{\rm Jup}$ planet may exist at  $\sim 100$--$120$~au. We also demonstrate that the ring seen in VLT/SPHERE observations at 96.7~au is consistent with the mass and semi-major axis of WISPIT 2b and confirm that WISPIT 2b cannot be the cause of the ALMA ring. We also discuss how future observations of the system's gas component may help constrain these predictions better, and note that this third planet in WISPIT 2 may be discoverable using upcoming JWST/NIRCam observations or {\em Gaia} data release 4 (DR4).


\end{abstract}



\section{Introduction}
\label{sec:intro}

Of the several hundred protoplanetary discs that have been imaged so far, only a handful have been confirmed to be hosting embedded planets. The most notable examples of these include PDS 70 \citep{Keppler2018, Haffert2019} and WISPIT 2 \citep{vanCapelleveen2025, Lawlor2026}. Both systems have two confirmed protoplanets producing observable ringed substructures in the disc. Although dust rings caused by planets are present in both systems, the rings themselves do not prove the existence of the planets. Indeed, there are a number of other processes by which such dust pileups can form including, but not limited to, rapid dust growth due to the condensation of volatiles at snow lines \citep{zhang2015}, material flows at the dead-zone edge \citep{flock2015}, self-organised zonal flows \citep{Bethune2016, Riols2018}, or inhomogeneities in the gas pressure \citep{pinilla2012}. 

In both PDS 70 and WISPIT 2, rings hinted at the possible presence of planets, but more rigorous metrics were needed to confirm the planets' existence. VLT instruments, such as MUSE \citep[PDS 70c,][]{Haffert2019}, SPHERE (PDS 70b and WISPIT 2b, detected by \citealp{Keppler2018} and \citealp{vanCapelleveen2025}, respectively), and infrared interferometry with VLTI/GRAVITY+ \citep[WISPIT 2c,][]{Lawlor2026} were used to image directly and hence, verify the existence, masses and locations of the previously hypothesised planets. A third planet has been hypothesised in both systems \citep[e.g.][]{Hammond2025, Facchini2026, Lawlor2026, Balsalobre-Ruza2026, Trevascus2026, vanderMarel2026} but these are yet to be detected.

\begin{figure}
    \centering
    \includegraphics[width=1\linewidth]{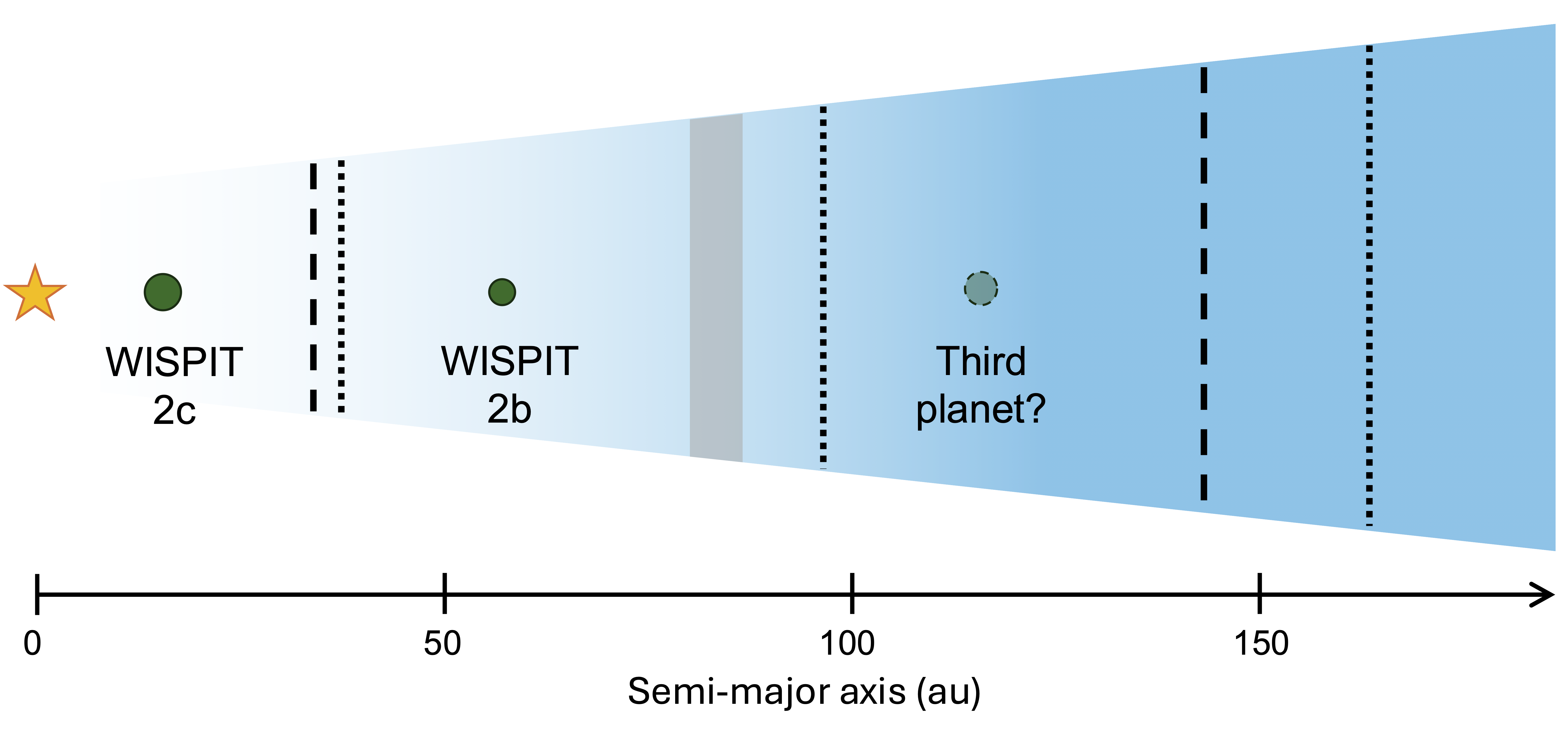}
    \caption{Schematic edge-on view of the inner part of the WISPIT 2 system. Dotted and dashed lines indicate the positions of the peaks of the rings detected using VLT/SPHERE and ALMA, respectively. The solid grey region marks the location of faint millimetre dust detected by \citet{vanderMarel2026}. The ALMA ring at 144.4~au sits unusually far from the outermost planet, WISPIT 2b, so may hint at the presence of a third planet between $\sim$100--140~au.}
    \label{fig:schematic}
\end{figure}

Figure \ref{fig:schematic} shows a schematic of the observed substructures and planets in the system.  The WISPIT 2 system comprises a spectroscopic binary \citep{Burgy2026} with a large ($\sim 380$~au) circumbinary disc \citep{vanCapelleveen2025}. The stars have masses of 0.97 and 0.33~$M_\odot$, with an orbital separation of 0.072~au (period of $4.8\pm0.1$~days). VLT/SPHERE observations show a complex, multi-ringed structure to the disc \citep{vanCapelleveen2025}. Rings have been observed in the near infrared at 38.4, 96.7, 163.6, and 316.4~au. Additionally, two rings have been detected in ALMA continuum images of the system, at 144.4~au \citep{Facchini2026} and 33.2~au \citep{Benisty2026}. Recent work by \citet{vanderMarel2026} also detected a faint region of millimetre dust at $\sim 80$--$84$~au.

The system hosts two confirmed planets: WISPIT 2b with a mass of $4.9^{+0.9}_{-0.6}~M_{\rm Jup}$ at an orbital semi-major axis (SMA) of 57.3~au \citep{vanCapelleveen2025}, and WISPIT 2c with an estimated mass of 8--$12~M_{\rm Jup}$ at a much closer orbit of $\sim$14~au \citep{Lawlor2026}. Although WISPIT 2b and 2c can explain some of the gaps and rings seen in the disc, \citet{Facchini2026} notes that the outermost planet, WISPIT 2b, sits at an unusually large distance from the ALMA ring exterior to its orbit. This may suggest that WISPIT 2b has a much higher mass than current estimates (approximately three times what it is currently believed to be, according to \citealp{Facchini2026}), allowing it to perturb the dust much further away. More novel interpretations can also be considered. For instance, WISPIT 2b may be undergoing intermittent migration, progressively speeding up and slowing down such that the ring it forms remains at its original location while the planet moves inwards \citep[e.g.][]{Wafflard2020, Weiskopf2026}. \citet{Facchini2026} also proposed that WISPIT 2b may have a large orbital eccentricity, pushing the inner edge of the resultant ring further away. However, orbital fitting by \citet{vanderMarel2026} shows that the eccentricity of WISPIT 2b ($\sim 0.12$) is inconsistent with the highly circular ring shape of the 144~au ring. The final possibility is that there may be an additional, undetected planet exterior to WISPIT 2b, causing the ALMA ring seen at 144.4~au \citep{Facchini2026, Lawlor2026}. This would also explain the discrepancy in the eccentricities of WISPIT 2b and the 144~au ring: an additional planet on a circular orbit may be producing the circular ring at 144~au \citep{vanderMarel2026}. \citet{vanderMarel2026} provide a preliminary estimate based on orbital fitting that such a planet would have a mass of $\sim 0.5~M_{\rm Jup}$  and \citet{Facchini2026} estimate its semi-major axis to be $\sim 130$~au based on the gas gap location. 

While direct imaging is widely considered the gold standard of discovery for planets embedded in protoplanetary discs, lower mass planets ($\lesssim M_{\rm Jup}$) can be difficult to detect in this way \citep{Asensio-Torres2021}. Using hydrodynamical simulations, \citet{Faruqi2026} derived a relationship between a ring's location and a planet's mass, which is independent of the dust size and hence, can be observed at any wavelength. Assuming that an additional planet exists in the WISPIT 2 system, this provides a constraint on the planet's mass and semi-major axis. Since the planet has not currently been detected, we can apply an additional constraint on the basis that the planet mass must fall below detection capabilities. We can also assume that the planet must be massive enough not only to form the observed substructures but to do so within the lifetime of the system.

In this article, we combine the methods described in \citet{Faruqi2026} with observational limits and the gap-opening timescale of \citet{Cordwell2024} to estimate the range of possible masses and semi-major axes of a potential third planet in WISPIT 2.


\section{Constraints on the mass and semi-major axis of the planet}
\label{sec:methods}

\subsection[Ring location prescription by A.\ Faruqi et al.\ 2026]{Ring location prescription by \citet{Faruqi2026}}

Given that neither the mass nor the orbital radius of the potential third  planet in WISPIT 2 are currently known, we apply the prescription by \citet{Faruqi2026}  to find the mass--SMA combinations that could form the ALMA ring. 

\citet{Faruqi2026} ran a series of 2D hydrodynamic simulations using \textsc{FARGO3D} \citep{masset2000fargo, llambay2016fargo} to explore the connection between observable characteristics of dust rings and the masses and orbital radii of planets producing them. They demonstrated that for all Stokes numbers and aspect ratios, the pressure peak location scales with approximately 5.3 times the Hill radius of the planet ($R_{\rm Hill}$).  Inverting that relation, the mass of an (unseen in this case) ring-forming planet can be related to the ring's observed peak location, $r_{\rm peak}$, according to the following relation:
\begin{equation}
\label{eq:ring_peak}
    \frac{M_{p}}{M_{*}} = 3 \left[\frac{1}{5.3} \left(\frac{r_{\mathrm{peak}}}{r_{p}}-1.07\right) \right]^{3}
\end{equation}
where $M_{p}$ is the planet mass, $M_{*}$ is the stellar mass, and $r_{p}$ is the planet's semi-major axis.

This equation was derived for discs containing only a single planet. However, we refer to simulations by \citet[][their Figure~2]{Pinilla2015a} and \citet[][their Figure~2]{Pritchard2024}, which feature two ring-forming planets within the same disc, to demonstrate that although the peak density in the ring differs when a second planet is present, the radial location of the ring peak does not. This expression also assumes a single central star but given the small separation of the WISPIT~2 spectroscopic binary (0.072~au, around 200~times tighter than the separation of the closest-in planet), it is reasonable to use the total binary mass for $M_{*}$. Thus, we apply equation \ref{eq:ring_peak} to both WISPIT 2b  and 2c (to determine where rings formed by these planets should reside) and the ALMA ring at 144.4~au (to constrain the combinations of planet masses and semi-major axes that could be causing it). 

\subsection{Gap-opening timescale}
\label{sec:gap_opening_timescale}

In WISPIT 2, multiple gaps have been identified in measurements of carbon monoxide (CO) isotopologues by \citet{Benisty2026} and \citet{vanderMarel2026}, as well as in scattered light data from VLT/SPHERE, which approximately trace the disc's gas component \citep{vanCapelleveen2025}. The most prominent gap is situated at approximately 69~au, with the known planet, WISPIT 2b, residing within its 59~au extent. An additional, shallower gap is also identified at approximately 130~au, which may be caused by an undetected planet \citep{Facchini2026, vanderMarel2026}.

The formation of a gas gap takes time, depending on the mass and semi-major axis of the perturbing planet. A gap-forming planet in the system must have sufficient mass not only to form a gap, but also to do so on timescales shorter than the age of the system \citep{Hourigan1984, Ward1989}. This consideration provides a second useful constraint on properties of the unseen planet.  The point is that if the planet has fully cleared its gap, we can rule out a region of the potential planet's mass--SMA parameter space by identifying the planet masses and planetary orbital radii for which the gap-opening timescale is greater than the age of the system, $5.1^{+2.4}_{-1.3}$~Myr \citep{vanCapelleveen2025}. 

\citet[][equation 32]{Cordwell2024} present a refined form of the gap-opening timescale from \citet{Hourigan1984} and \citet{Ward1989}. They define the gap-opening timescale, $\tau_{\rm gap}$, as:
\begin{equation}
\label{eq:tgap}
    \tau_{\rm gap} = \frac{P_{p}}{2} \left( \frac{h_{p}}{r_{p}} \right)^{-1} \left( \frac{M_{p}}{M_{\rm th}}\right)^{-2} B
\end{equation}
where $P_{p}$ is the planet's orbital period, $h_{p}/r_{p}$ is the disc's aspect ratio at the planet's location, $M_{\rm th}$ is the thermal mass \citep{Goodman2001}, and
\begin{equation}
    B = 27.9 + 1.41 \left( \frac{M_{p}}{M_{\rm th}} \right)^{-1.78} .
\end{equation}
This is the characteristic timescale for a planet to deplete the local gas surface-density to approximately zero. We use this expression to compute the gap-opening timescale as a function of the planet's mass and semi-major axis to apply an approximate limit to the range of possible masses and orbital radii for the unseen, gap-forming planet. A partially depleted gap could form on shorter timescales, and an observable partial gap could be produced by a lower-mass planet than equation \ref{eq:tgap} predicts. Therefore, the reliability of this constraint is subject to the degree of gas depletion in WISPIT 2.

For now, we assume that the gap seen at $\sim 130$~au is the full extent to which the unseen planet can clear gas from the region of its orbit. The relationship between a partial gap's depth and planet mass is discussed in Section \ref{sec:discussion_gaps}. We use the gas aspect ratio defined by the parameters in Supplementary Table 5 in \citet{vanderMarel2026} and calculate the gap-opening timescale as a function of planet mass and semi-major axis. 

\subsection{Observational Limits}
\label{sec:obs_limits}
WISPIT 2b was first inferred to exist by \citet{Close2025}, using H$\alpha$ emission to measure the accretion rate at the planet's location and concluding that this must correspond to an accreting protoplanet. High resolution VLT/SPHERE scattered light images of WISPIT 2 were later presented in \citet{vanCapelleveen2025}, in which WISPIT 2b was detectable. We argue that for a planet to have remained undetected in WISPIT 2, it must fall outside the detection capabilities of  instruments that have previously been used to observe the system.

\citet{Asensio-Torres2021} discuss the detection limits of VLT/SPHERE for direct imaging of embedded protoplanets. They consider different models, based on different protoplanet formation mechanisms. Planets formed via faster mechanisms, such as disc instabilities or hot core accretion processes, are expected to form hotter \citep[e.g.,][]{Mordasini2012II, Bussmann2026}, making them more luminous and easier to detect \citep[e.g.,][]{Marley2007, Spiegel2012}. Alternatively, planets formed via slow core accretion might be somewhat cooler, and therefore harder to detect. However, the differences are minimal \citep{Mordasini2017}.

\begin{figure*}
    \centering
    \includegraphics[width=0.9\linewidth]{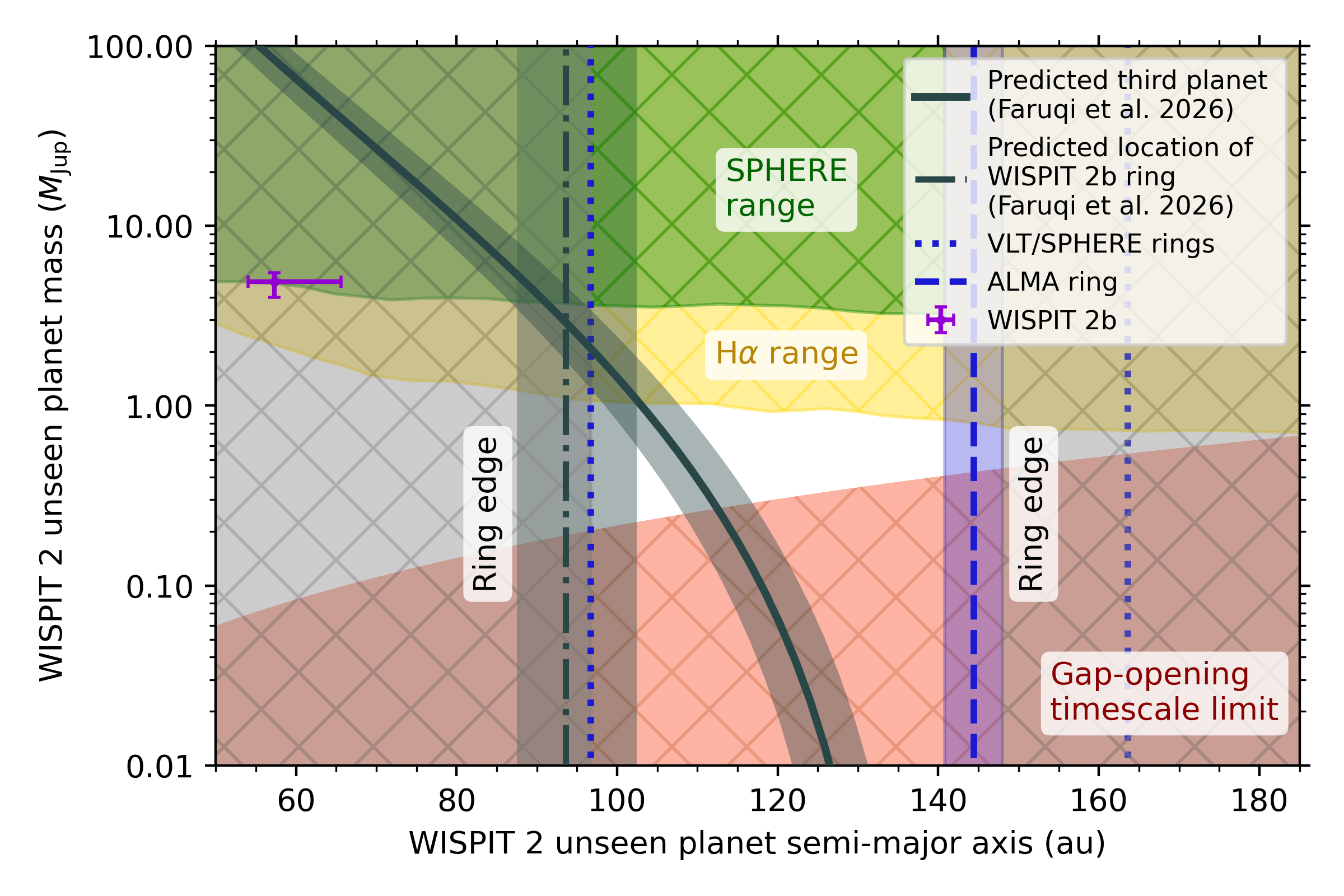}
    \caption{Mass--SMA parameter space for a potential, unseen planet in WISPIT 2. The solid dark grey line marks the range of possible masses and semi-major axes for the planet according to \citet[][equation 11]{Faruqi2026}, with errors denoted by the surrounding shaded region. The dotted and dashed lines mark the central locations of the observed SPHERE and ALMA rings, respectively. The dark grey dot-dashed line indicates the predicted location of a ring produced by WISPIT 2b, which agrees with the observed SPHERE ring. The hatched regions indicate the ``forbidden zones'' i.e., the regions of the parameter space we do not predict a planet can exist due to observational constraints (the SPHERE and MagAO-X H$\alpha$ detection ranges), the presence of rings, or the limit due to the gap-opening timescale exceeding the 5.1~Myr age of the system.}
    \label{fig:planet_mass_location}
\end{figure*}

We assume that the unseen planet in WISPIT 2 formed via a ``hot-start'' mechanism since warm or hot starts are considered to be much more likely than cold starts \citep[e.g.][]{Trevascus2025}. We take the data from the \textsc{AMES-DUSTY} model presented in Figure~2 of \citet{Asensio-Torres2021} and apply the limit it represents to our own parameter space (shown in green in Figure \ref{fig:planet_mass_location}).

We also consider the detection limits of H$\alpha$ emission from an accreting protoplanet detected using MagAO-X, the instrument used by \citet{Close2025} to discover WISPIT 2b. We use the 5$\sigma$ contrast curves shown in Figure~3 of \citet{Cugno2023} to obtain the minimum detectable H$\alpha$ flux for a planet in WISPIT 2 as a function of the planet's semi-major axis. This is then converted to a luminosity, assuming an extinction of $A_{\nu} = 3$ mag, as was done by \citet{Close2025} for WISPIT 2b. Extinction in WISPIT 2 has not yet been measured but \citet{Close2025} argue that, given the minimal interstellar extinction to the system and the clearing of material around WISPIT 2b, $A_{\rm V} = 0$--3~mag is a reasonable range. The resulting luminosity is then converted to an accretion luminosity, $L_{\rm acc}$, via the relation given in Equation~D.1 and Table~D.1 in \citet{Fiorellino2025} for H$\alpha$ emission. We assume a mass accretion rate, $\dot M_{\rm acc}$, which is approximately equal to the unseen planet's mass, $M_{p}$, divided by the age of the system, $\tau$. This is an order-of-magnitude estimate, based on the assumption that accretion has been occurring for approximately as long as the disc has existed. Studies on the ``formation delay'' suggest that a more realistic accretion rate would only be a factor of a few different from this \citep[e.g.][]{Fortney2005, Zhang2024}. We  then compute the mass detection limit for H$\alpha$ emission using the expression given in \citet{Gullbring1998}:
\begin{equation}
    M_{p} = \Bigg[ \bigg(1 - \frac{R_{p}}{R_{\rm in}} \bigg)^{-1} \frac{L_{\rm acc} R_{p} \tau}{G} \Bigg]^{1/2}
\end{equation}
where $R_{p}$ is the planet's radius and $R_{\rm in}$ is the inner radius of the planet's accretion disc. These values are taken to be $2~R_{\rm Jup}$ and $10~R_{\rm Jup}$, respectively, which are standard assumptions \citep[e.g.][]{Spiegel2012, Zhu2015, Cugno2023}. The expression by \citet{Gullbring1998} assumes magnetospheric accretion in stars, but has been used for planetary accretion as well \citep[e.g.][]{Zhu2015, Cugno2023}. The true accretion rate for a planet may be lower than the estimate $\dot M_{\rm acc}\sim M_p/\tau$, especially at a given epoch, so that the minimum planet mass detectable in H$\alpha$ emission may be somewhat higher than the estimate presented here, i.e., the non-detection less constraining.

\section{Results}

\subsection[Application of Faruqi et al. (2026) to WISPIT 2b and 2c]{Application of \citet{Faruqi2026} to WISPIT 2b and 2c}
We first test our proposed method on the two confirmed planets in this system. Figure \ref{fig:planet_mass_location} shows that, according to equation~\ref{eq:ring_peak}, the combination of the mass and the semi-major axis of WISPIT 2b (marked in purple) is highly inconsistent with the ALMA ring seen at 144.4~au (blue dashed line), as noted by \citet{vanCapelleveen2025}, but instead predicts that the ring formed by WISPIT 2b, the should be at at $93.6^{+8.8}_{-6.0}$~au (dark grey dash-dotted line). This aligns extremely well with the location of the ring seen in VLT/SPHERE images at 96.7~au, which is shown by the dotted blue line on the plot.  This success strongly motivates exploring the possibility of an additional, undetected planet in the disc, that is responsible for the ALMA ring at 144.4~au.\\

We also apply our method on WISPIT 2c and calculate a ring location of $26^{+5}_{-4}$~au. This differs slightly from the measured peak location of the ring exterior to WISPIT 2c, which has been measured at $\sim 33$--$38$~au \citep{vanCapelleveen2025, Benisty2026}, but does fall within the ring's extent, given its width of $8.9$~au \citep{Benisty2026}. This discrepancy results from the fact that WISPIT 2c, with a mass of $\sim 8$--$12~M_{\rm Jup}$, is well above the mass range used to derive equation \ref{eq:ring_peak} \citep{Faruqi2026}. In \citet{pinilla2012:a}, the authors also observed a similar linear relationship between a planet's Hill radius and associated ring location, but found that for planet masses beyond $\sim 5~M_{\rm Jup}$, the gradient increased to $\sim 10R_{\rm Hill}$. If we assume a gradient of $10R_{\rm Hill}$ (rather than $5.3R_{\rm Hill}$, as determined by \citealp{Faruqi2026}) and repeat our calculation with this modified form of equation \ref{eq:ring_peak}, we predict the ring to be located at $\sim 34.4$~au, which is consistent with observations. However, we caution that further modelling work is needed to accurately describe how the relationship described by equation \ref{eq:ring_peak} changes at higher planet masses.

\subsection{An unseen third planet?}

In Figure~\ref{fig:planet_mass_location}, we also present the result of combining all the constraints described in Section~\ref{sec:methods}. The solid dark grey line and associated error bars shown by the band surrounding it show our predicted mass--SMA combinations based on equation \ref{eq:ring_peak}. We do not expect the planet to reside beyond $\sim 120$~au due to the gap-opening timescale constraint (Section \ref{sec:gap_opening_timescale}), unless the gap is not fully formed (discussed further in Section \ref{sec:discussion_gaps}). Based on the observational constraints (Section \ref{sec:obs_limits}) coupled with the location of the SPHERE ring at $96.7$~au, the unseen planet is expected to reside beyond $\sim 97$~au and to have a mass less than $\sim 1~M_{\rm Jup}$. 

Therefore, we arrive at a limited range of possible mass and semi-major axis combinations for the planet, which span from $\sim 0.2$--$1~M_{\rm Jup}$ and $\sim 100-120$~au, respectively. A planet within this mass--SMA range would not have been detected in previous surveys. Planet masses below 0.2 $M_{\rm Jup}$ within this location range are possible but would only be capable of forming a partially depleted gas gap in the system. 

Our mass range is consistent with the mass estimate of $\sim 0.5~M_{\rm Jup}$ suggested by \citet{vanderMarel2026}, who used thermo-chemical modelling to approximate the planet mass that would be high enough to form the ALMA ring at 144.4~au and low enough that its associated gas gap is optically thick in $\rm ^{12}CO$. \citet{Facchini2026} and \citet{vanderMarel2026} both hypothesise the planet's semi-major axis to be $\sim 130$~au, coincident with the approximate mid-point of the shallow gap seen in VLT/SPHERE images by \citet{vanCapelleveen2025}. However, it is known that gap-opening planets do not necessarily reside at the exact radial centres of the gaps they produce \citep[e.g.][]{Durmann2015, Dipierro2016, meru2019ring}. \citet{pinilla2012} and \citet{rosotti2016minimum} found that rings are typically expected to form at around $1.1$--$1.3$ times the planet's orbital radius. For a ring at 144~au, this would correspond to a planet at $\sim 110$--$130$~au. The width of this gap in the gas has not been directly measured due to the optically thick nature of  $\rm ^{12}CO$ line emission. However, \citet{vanderMarel2026} model the gap to extend from 117 to 143~au (assuming it extends $2R_{\rm Hill}$ in either direction from the gap centre), making it plausible that the unseen planet's location falls within our predicted range, while still being consistent with the measured location of the gas gap.

\section{Discussion}
\label{sec:discussion}
\subsection{Does WISPIT 2 have an unseen planet?}
Not every ring has to correspond to a planet. There are mechanisms by which a single planet can form multiple rings. For instance, in systems with low turbulent viscosities, a single planet can produce multiple rings \citep[e.g.][]{Dong2017, Fedele2018}. However, in \citet{Dong2017} and \citet{Fedele2018}, the additional rings lie interior to the planet's orbit, whereas in WISPIT 2, additional rings exist exterior to the outermost known planet. 

We conclude that a currently undiscovered planet in the WISPIT 2 system, located at $\sim 100$--$120$~au with a mass of $\sim 0.2$--$1~M_{\rm Jup}$, is consistent with the locations of the observed ALMA and SPHERE rings, while also being undetectable by prior observations of the system. WISPIT 2b, even if three times more massive than expected \citep{Facchini2026} is, according to our calculations, highly unlikely to be capable of forming the ring seen at 144.4~au.




\subsection{What can we learn from the system's gaps?}
\label{sec:discussion_gaps}
\begin{figure}
    \centering
    \includegraphics[width=1\linewidth]{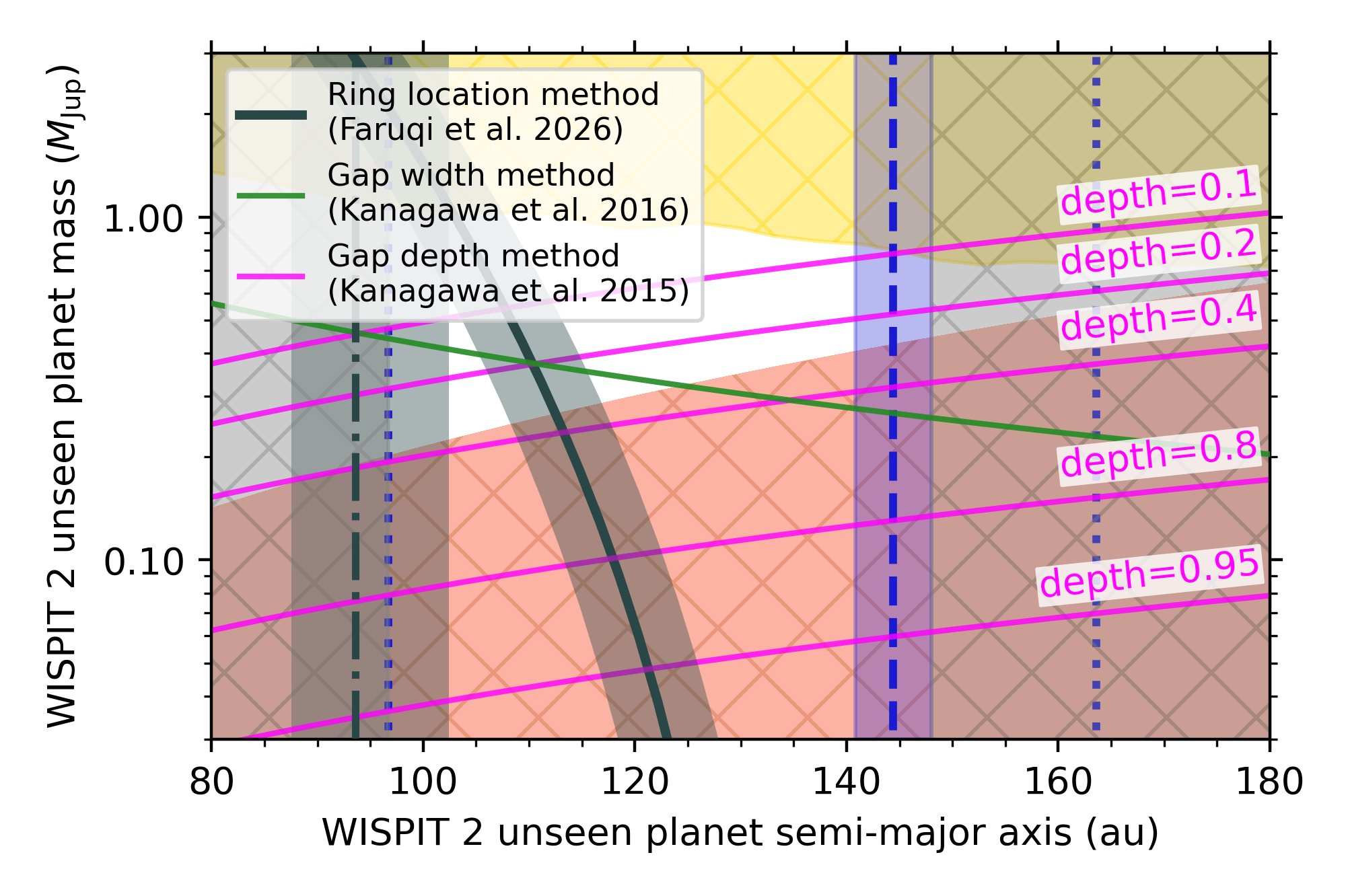}
    \caption{Zoomed in version of Figure \ref{fig:planet_mass_location} but with the \citet{Kanagawa2015,Kanagawa2016} gap and depth prescriptions overlaid. Since the depth of the gas gap has not been directly measured, we plot the depth relation for a number of different values of $\Sigma_{p}/\Sigma_{0}$ and find that all three prescriptions intersect at a depth of 0.2. }
    \label{fig:Kanagawa}
\end{figure}

\citet{vanderMarel2026} recently used DALI thermo-chemical models to constrain the depth of the gas gap associated with the hypothesised, unseen planet and found that their best fit model predicted a gap depth of approximately 0.95 (i.e. a 5\% drop in the gas surface density at the planet's location). 


Two widely used methods for calculating planet properties via observable substructures in discs focus on the gap depth and gap width prescriptions provided by \citet{Kanagawa2015} and \citet{Kanagawa2016}, respectively. These relate the mass and semi-major axis of a planet to the observable properties of disc substructure, in this case, the gas gap depth and width.  How does this approach align with our complementary focus on ring properties?

The \citet{Kanagawa2015} depth prescription is based on a detailed analytic calculation of the angular momentum exchange between a protoplanetary disc and an embedded planet. They define $K$, which relates the depth of a disc's gas gap to the mass and semi-major axis of the planet producing it:
\begin{equation}
\label{eq:Kdepth}
    K = \Bigg ( \frac{M_{p}}{M_{*}} \Bigg )^{2} \Bigg (\frac{h_{p}}{r_{p}} \Bigg )^{-5} \alpha^{-1}
\end{equation}
where $\alpha$ is the Shakura--Sunyaev viscosity parameter \citep{shakura1973} and all other parameters are as previously defined. $K$ is then related to the fractional gap depth ($\Sigma_{p}/\Sigma_{0}$) by the following relation:
\begin{equation}
\label{eq:gap_depth}
    \frac{\Sigma_{p}}{\Sigma_{0}} = \frac{1}{1+0.04K}.
\end{equation}

The \citet{Kanagawa2016} width prescription is based on empirical data from a series of hydrodynamical models run using \textsc{FARGO} \citep{masset2000fargo}. This defines a new parameter, $K'$, as
\begin{equation}
\label{eq:Kwidth}
    K' = \Bigg ( \frac{M_{p}}{M_{*}} \Bigg )^{2} \Bigg (\frac{h_{p}}{r_{p}} \Bigg )^{-3} \alpha^{-1}.
\end{equation}
The relationship between gap width, $\Delta_{\rm gap}$, and $K'$ is
\begin{equation}
\label{eq:gap_width}
    \frac{\Delta_{\rm gap}}{r_{p}} = 0.41 K'^{1/4}.
\end{equation}

Currently, neither the width nor the depth of the gas gap at $\sim 130$~au has been directly measured. However, detailed modelling by \citet{vanderMarel2026} has provided a best fit for the gap depth of $\Sigma_{p}/\Sigma_{0}=0.95$ and an estimate for the gap width of $\sim 26$~au ($4R_{\rm Hill}$). Using these values, we can approximate the range of possible masses and semi-major axes for the unseen planet in WISPIT 2.

We take the gas component's aspect ratio to be $h_{p}/r_{p} = 0.12(r_{p}/150)^{0.5}$, as defined in \citep{vanderMarel2026}. We also assume the disc's $\alpha$-viscosity to be $10^{-4}$ \citep[e.g.][]{Lesur2023}, since there are currently no constraints on the system's viscosity. We can then use equations~\ref{eq:gap_depth} and~\ref{eq:gap_width} to obtain a rough estimate of the planet masses and semi-major axes that would be consistent with the works of \citet{Kanagawa2015} and \citet{Kanagawa2016}, while also satisfying the criteria outlined in Section~\ref{sec:methods}. 

In Figure \ref{fig:Kanagawa}, we plot the expected planet mass as a function of the potential planet's semi-major axis, based on the gap width and depth. We trial a few different gap depths and find that for a gap depth of 0.2, the Kanagawa prescriptions intersect with the ring location prescription at a semi-major axis and mass of approximately 110~au and $0.5~M_{\rm Jup}$, respectively. However, a gap depth of 0.2 is inconsistent with the gap depth predicted by the DALI models presented in \citet{vanderMarel2026}. Their predicted gap depth of 0.95, intersects with our ring location prescription at a mass of $\sim 0.05~M_{\rm Jup}$ and a semi-major axis of $\sim 123$~au, and does not intersect with the estimated gap width of 26~au \citep{vanderMarel2026} at all. Interestingly, a planet at $\sim 123$~au would be in a 3:1 Mean Motion Resonance (MMR) with WISPIT 2b, as noted by \citet{Facchini2026}. The range of mass and locations predicted by equations \ref{eq:Kdepth} and \ref{eq:gap_depth} for a gap depth of 0.95 also all fall below the gap-opening timescale threshold, suggesting that if the unseen planet is causing this gap, it is still in the process of forming.






\subsection{How could this planet be detected?}
\label{sec:how_to_detect_wispit2d}
\citet{Lin2026} demonstrate that JWST/NIRCam has the potential to detect planets embedded in a protoplanetary disc and quantify the detection limits for the TW Hya system. We adapt their contrast curves \citep[see Figure 11 in][]{Lin2026} to the distance of WISPIT 2 and account for its lower (approximately half) stellar flux relative to TW Hya to estimate the minimum detectable planet mass. We find that in the absence of local extinction, JWST/NIRCam observations of WISPIT 2 should be sensitive to planets with masses as low as approximately $0.2~M_{\rm Jup}$ at $\sim 110$~au in WISPIT 2. Given that we expect the planet mass to be 0.2--1~$M_{\rm Jup}$ (assuming it is fully clearing a gap), this should be detectable. JWST observations of WISPIT 2 are scheduled for late 2027, including resolved spectroscopy and coronagraphic imaging, which may reveal the presence of the additional planet.

Furthermore, the upcoming  {\em Gaia} data release (DR4) may enable the astrometric detection of a planet in WISPIT 2. Stellar astrometric observations have been shown to detect and characterise the masses and orbital properties of long-period sub-stellar and planetary companions with recent {\em Gaia} data releases (DR2 and DR3, \citealt{Gaia2023}). Such studies can also place constraints on the presence of planets in disc-hosting systems \citep{Vioque2026}. The upcoming {\em Gaia} DR4 will report a catalogue of planetary candidates as well as ``per epoch'' astrometric data, allowing for detailed planetary orbit fitting to be conducted with 1900$\pm$540 new giant, long-period planets expected to be found \citep{Lammers2026}, including those orbiting binary stars \citep{Baycroft2026}.

We follow the methodology of \citealt{Lammers2026} to simulate per-epoch astrometry for WISPIT~2 using known stellar properties \citep{vanCapelleveen2025, Close2025, Burgy2026, Lawlor2026} and predicted {\em Gaia} observations from the {\em Gaia} Observing Forecast Tool \citep{Holl2023}. Using {\sc gaiamock} \citep{ElBadry2024}, we generate a 5.5-yr astrometric time series spanning the {\it Gaia} DR4 baseline while varying the planet's mass and semi-major axis as per the expectations from Figure \ref{fig:planet_mass_location} ($a = 96$--102~au, $M_{\rm p} = 0.6$--1.0~$M_{\rm Jup}$), and fit these data with {\sc emcee} \citep{Foreman-Mackey2013}. 

Although the planet's long period prevents reliable orbital characterisation due to only a short segment of its orbit being traced during the observing period of DR4, its predicted along-scan displacement ($\sim$0.15\,mas) lies above the {\em Gaia} astrometric noise floor ($\sim$0.05\,mas) for the {\it Gaia} magnitude of WISPIT 2 \citep{Holl2023, ElBadry2024}. This means that should this planet exist, {\em Gaia} DR4 may have the precision to detect it if the inner binary and other planetary companions are accurately modelled.

\section{Conclusions}

In this article, by applying the results of our ring based method for determining planetary properties \citep{Faruqi2026}, we successfully predict the association of WISPIT 2b with the ring at 96.7~au. Based on this, we restrict the range of possible masses and semi-major axes of a third planet in the system, which we argue could be responsible for forming the 144.4~au ring.   We predict that the unseen planet would have a mass of $\sim 0.2$-$1~M_{\rm Jup}$ and an orbital radius of $\sim 100$--120~au. If the planet is still in the process of forming, these limits are less stringent and planet masses down to $\sim 0.01~M_{\rm Jup}$ and semi-major axes out to $\sim 130$~au could be possible, while still forming a shallow, partial gap. 

We discuss how upcoming observations may reveal our predicted planet. JWST may be able to detect a planet of our predicted mass and semi-major axis within the next year. A more direct approach is made possible by the upcoming {\em Gaia} data release (DR4): we find that our predicted planet may produce a detectable astrometric signal. 


\begin{acknowledgments}
The authors thank the referee for helpful suggestions that improved the manuscript. The authors thank Jessica Speedie for insightful conversations on observations of WISPIT 2. AF acknowledges support from the Royal Society Enhancement Award. FM acknowledges support from The Royal Society Dorothy Hodgkin Fellowship. This project has received funding from the European Research Council (ERC) under the European Union's Horizon 2020 research and innovation programme (Grant agreement No.~101020057). PN is supported by a Warwick Prize Fellowship. G-DM acknowledges the support of the Deutsche Forschungsgemeinschaft (DFG) through grant MA~9185/2-1.  REP is supported by a Discovery grant from the National Science and Engineering Council (NSERC) of Canada. TGW acknowledges support from the UKSA and University of Warwick.
\end{acknowledgments}

\bibliography{bibliography}{}
\bibliographystyle{yahapj} 



\end{document}